\documentclass[12pt]{article}
\usepackage{amsmath,amssymb,url,graphicx,dcolumn,bm}
\usepackage[usenames,dvipsnames]{color}

\title{Is the Hypothesis About a Low Entropy Initial State of the Universe Necessary for Explaining the Arrow of Time?}

\author{
Sheldon Goldstein\footnote{Departments of Mathematics, Physics, and Philosophy, 
	Rutgers University, Hill Center, 110 Frelinghuysen Road, 
	Piscataway, NJ 08854-8019, USA. Email: oldstein@math.rutgers.edu},\  
Roderich Tumulka\footnote{Department of Mathematics, 
	Rutgers University, Hill Center, 110 Frelinghuysen Road, 
	Piscataway, NJ 08854-8019, USA. Email: tumulka@math.rutgers.edu},\ and
Nino Zangh\`{\i}\footnote{Dipartimento di Fisica, Universit\`a di Genova, and
	Istituto Nazionale di Fisica Nucleare (Sezione di Genova),
	Via Dodecaneso 33, 16146 Genova, Italy. Email: zanghi@ge.infn.it}
}

\date{July 6, 2016}

\newcommand{\be}{\begin{equation}}
\newcommand{\ee}{\end{equation}}
\newcommand{\vv}{\boldsymbol{v}}
\newcommand{\RRR}{\mathbb{R}}
\newcommand{\SSS}{\mathbb{S}}
\newcommand{\vq}{\boldsymbol{q}}
\newcommand{\vp}{\boldsymbol{p}}
\newcommand{\cm}{\mathrm{cm}}

\begin{document}
\maketitle
\begin{abstract}
According to statistical mechanics, 
micro-states of an isolated physical system (say, a gas in a box) at time
$t_0$ in a given macro-state of less-than-maximal entropy typically 
evolve in such a way that the entropy at time $t$ increases with $|t-t_0|$ in \emph{both} time
directions. In order to account for the observed entropy increase in only
\emph{one} time direction, the thermodynamic arrow of time, one usually
appeals to the hypothesis that the initial state of the universe was one of very low entropy. In certain recent models of cosmology,   however, no
hypothesis about the initial state of the universe is invoked. We 
discuss how the emergence of a thermodynamic arrow of time 
in such models can nevertheless be compatible with
the above-mentioned consequence of statistical mechanics, appearances to the contrary notwithstanding.

\medskip

\end{abstract}

\section{Background}

The goal of this paper is to discuss an explanation of the ultimate origin of the thermodynamic arrow of time that was recently proposed by Carrol and Chen \cite{CC04,CC05,Car} and later explored also by Barbour, Koslowski, and Mercati \cite{BKM13,BKM14,BKM15}. The cosmological models proposed by these authors undergo an entropy increase as demanded by the second law of thermodynamics but do not postulate a special low-entropy initial state of the universe. As we explain, such a behavior would appear to be excluded by standard facts of statistical mechanics, but, as we also explain, in fact it is not. We begin with describing the problem that these models address.

Why should there be an arrow of time in our universe, 
governed as it is, at the fundamental level, by reversible microscopic laws?  Part of the answer is that a system in a state of low entropy, corresponding classically to small phase space volume, will tend to evolve to a (more likely) state of higher entropy, corresponding to much larger volume. This leads to the question, What is 
the origin of the low entropy initial states? If they are so unlikely, why should 
systems find themselves in such states?   In many cases, the answer is that we created them from states of lower entropy still. If we continue 
to ask such questions, we come to the conclusion that the cause of low entropy 
states on earth is 
a cosmological low entropy state in the distant past for the universe as a whole; see, e.g., \cite[Chap.~7]{Pen}, \cite[II \S 87]{Bol1898}, \cite{L07,ehrenfest,Davies,Pen79,HPMZ94}.

And what about the origin of this state? 
Penrose \cite{Pen} estimated the volume of the region of phase space corresponding to the possible initial states of the universe to be one part in $10^{10^{123}}$ of the entire relevant phase space. 
Why should the universe have begun in such an exceedingly improbable 
macro-state? 

An answer that has often been suggested, starting with Boltzmann \cite[II \S 90]{Bol1898},  is that such a state arose 
from a fluctuation out of equilibrium. In fact, if the universal dynamics were 
ergodic, such a fluctuation would  eventually occur, repeatedly, for all phase points 
with the possible exception of a set of measure $0$. 
Nonetheless, this {\em fluctuation scenario} is quite unsatisfactory; indeed, according to Feynman \cite[p.~115]{Fey} it is ``ridiculous.'' The problem is that if the explanation of entropy 
increase and the arrow of time in our universe is that they have emerged from 
a low entropy state that arose from a fluctuation, then that fluctuation should 
have been no larger than necessary---that is, to a state like the present state of 
the universe, and not to a state of much lower entropy as seems to have existed in 
the past. 
In the words of Eddington \cite{Edd}: 
``A universe containing mathematical physicists 
will at any assigned date [in a fluctuation scenario] be in the state of maximum disorganization which
is not inconsistent with the existence of such creatures.''
Here is a quotation from Feynman, referring to astronomy, to history books and history, 
and to paleontology: ``Since we always make the prediction that in a place where we have not looked we shall see stars in a similar condition, or find the same statement about Napoleon, or that we shall see bones like the bones that we have 
seen before, the success of all those sciences indicates that the world 
did not come from a fluctuation \ldots Therefore I think it is necessary to 
add to the physical laws \emph{the hypothesis that in the past the universe 
was more ordered \ldots than it is today}---I think this is the additional 
statement that is needed to make sense, and to make an understanding 
of the irreversibility'' \cite[p.~115, emphasis added]{Fey}.
Henceforth, adopting the terminology of Albert \cite{Alb}, we shall refer to what  has been  italicized, an idea originally due to Boltzmann, as the {\em past hypothesis}: that at the Big Bang, the universe was in an appropriate low-entropy macro-state.
(Albert meant a bit more than this by the past hypothesis. He meant that the initial micro-state of the universe should be regarded as uniformly distributed in the appropriate low-entropy macro-state.)

\section{Arrow of Time Without the Past Hypothesis}

That the past hypothesis, or some hypothesis on universal initial conditions \cite{Pen79}, is needed to account for the origin of the arrow of time has long been the received view. We discuss here an alternative view developed by Carroll and Chen \cite{CC04,CC05,Car} and explored also by Barbour et al.~\cite{BKM13,BKM14,BKM15}. Carroll and Chen proposed a cosmological model in which (what is usually called) the Big Bang does not involve a space-time singularity and is not the beginning of space-time; in fact, in their model time extends infinitely into the past. As a consequence, the time of the Big Bang is not the initial time, and to add to the physical laws a condition on the state at that time seems unnatural. Accordingly, Carroll and Chen (and Barbour et al.)\ do not postulate a past hypothesis. They claim that its consequences can nonetheless be derived from the dynamical laws and the typical behavior of the solutions. If that is correct, such an explanation of the arrow of time may be attractive because it avoids the postulation of an extremely unlikely state.

Carroll and Chen's claim that they do not need the past hypothesis is surprising. 
After all, if one assumes that the universe is governed by equations of motion that are reversible (and volume preserving, in the classical case, or unitary, in the quantum case) and one does not invoke anything like the past hypothesis, then it would seem that the probabilistic reasoning that leads us to expect that entropy increases towards the future, given the evidence of the present macro-state, would lead us also to expect the same increase toward the past. In a similar vein, it would seem that without the past hypothesis, a typical solution of the equations of motion should describe a universe that is in global thermal equilibrium throughout its history (except for fluctuations) \cite{ehrenfest,LPSW08}, in contrast to our universe. It would seem, in other words, that without the past hypothesis we would be stuck with the fluctuation scenario discussed above.

Notwithstanding the above reasoning, Carroll and Chen 
have proposed a cosmological model without any past hypothesis which has, nevertheless, an arrow of time. In fact, in this model, and in that of Barbour et al., the following is true:

\medskip

\noindent{\bf Fact 1.} The entropy $S(t)$ increases monotonically with $t$ (except for fluctuations) for $t>\tau$ and decreases monotonically with $t$ for $t<\tau$ from some ``central time'' $\tau$; thus, the (thermodynamic) arrow of time always points away from $\tau$. 

\medskip

A key ingredient of this model is that the volume of (the relevant) physical 3-space grows unboundedly as $t\to\pm\infty$, with matter distributed in it in such a way  that there is no upper bound for the entropy, 
and thermal equilibrium is never attained. This may explain, at least in part, why the model universe is not in global thermal equilibrium throughout its history. But there is also the following standard fact (see, e.g., \cite[II \S 88]{Bol1898}, \cite{L07,ehrenfest,HPMZ94,LPSW08,Lan}) of the statistical mechanics of a closed system:

\medskip

\noindent{\bf Fact 2.}  Most micro-states in a macro-state with less-than-maximal entropy $S_0$ at time $t_0$ (for example the present) evolve so that $S(t)$ increases with $|t-t_0|$ in both time directions.

\medskip

And, since $t_0$ in Fact 2 is different from $\tau$ in Fact 1, the following problem remains: How can Fact 2 be reconciled, without the invocation of a past hypothesis,  with the existence of a time $\tau$, significantly before $t_0$, such that the entropy $S(t)$ monotonically increases with $t$ for $t>\tau$?  We address this question by means of a simple toy model (also due to Carroll) that has all features relevant to our purposes.

In the next section we describe the toy model. In the rest of the paper we elaborate on the conflict between Fact 1 and Fact 2. We first describe and resolve an apparent mathematical contradiction associated with the Facts. We then formulate a more detailed argument that Fact 2 shows that the past hypothesis cannot be avoided, after which we describe what is wrong with that argument.

\section{Toy Model}

Consider a gas of $N\gg 1$ non-interacting classical particles moving freely in 3-dimensional Euclidean space. That is, the position of the $i$-th particle ($i=1,\ldots,N$) at time $t$ is given by
\begin{equation}
\vq_i(t) = \vq_i(0)+t\vp_i(0);
\end{equation}
the particle's momentum is $\vp_i(t)=\vp_i(0)$, and its mass is $m_i=1$. As $t\to\pm\infty$, the particles will be dispersed over larger and larger regions.
We  shall work in the center-of-mass Galilean frame, in which at time $0$ and thus at any $t$, $\vq_\cm = N^{-1} \sum_{i=1}^N \vq_i=0$ and $\vp_\cm = \sum_{i} \vp_i=0$.

\begin{figure}
	\centering
        \includegraphics[width=0.60\textwidth]{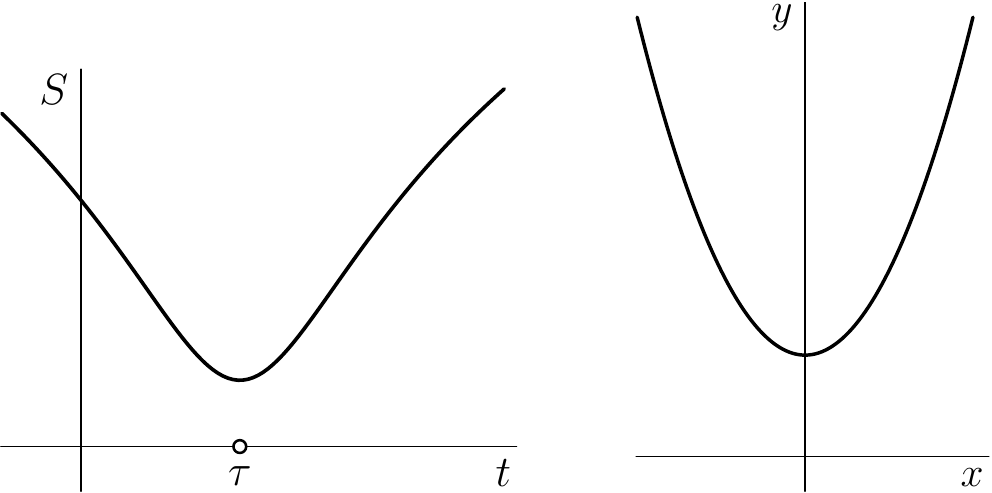}
	\caption{
	Left: entropy of the gas in the toy model as a function of time. 
	Right: orbit of the system in the reduced (2-dimensional) 
	phase space of the system, described in the text.}
	\label{fig:curve}
\end{figure}

As a quantitative measure of the size of the region
in 3-space occupied by the gas, 
we use the total moment of inertia\footnote{Actually, the total moment of inertia (i.e., the trace of the moment of inertia tensor) equals $2Ny$.}
\begin{equation}  \label{sigmadef}
y= N^{-1} \sum_{i=1}^N  {\vq_i}^2.
\end{equation}
It follows that $y(t)$ is a quadratic function of $t$, in fact
\begin{equation}\label{sigmat}
y(t) = 2EN^{-1} \bigl(t-\tau\bigr)^2 + \alpha
\end{equation}
with $E=\tfrac{1}{2} \sum_{i} \vp_i^2$ the kinetic energy (= total energy) and $\tau$ and $\alpha\ge 0$ suitable constants. From \eqref{sigmat} we see that,  $y(t)\to\infty$ as $t\to\pm\infty$, and, moreover, $y(t)$ assumes a unique minimum at the \emph{central time} $\tau$.
 
Now regard $E$, $y$, and $\tau$ as functions on phase space $\mathbb{R}^{6N}=\{(q,p)\}=\{(\vq_1,\ldots,\vq_N,\vp_1,\ldots,\vp_N)\}$. We denote by $\Gamma_E$ the set of points in phase space with  energy $E$ and $\vp_\cm=\vq_\cm=0$. The quantity
\begin{equation}\label{x}
 x(q,p)=-\tau(q,p)
\end{equation}
is the time that has elapsed since the central time if the system is presently in the state $(q,p)$.
Using the macro-variable $y$ to define the macro-states, the (Boltzmann) entropy of the gas is\footnote{The choice of $y$ as the sole macro-variable is an over-simplification for the sake of simplicity of the example. If we adopted the choice of macro-variables from the kinetic theory of gases (and thus the definition of Boltzmann entropy as $-k\int f \log f \, d^3\vq\, d^3\vp$ with $f$ the coarse-grained empirical density in the 1-particle phase space), then entropy would never change in the toy model \cite{Rei12}.}
\begin{equation}\label{entropy1} 
S(q,p)= k \log  \rho_{E(q,p)}( \Gamma_{y_0})\,.
\end{equation}
Here, $k$ is Boltzmann's constant, $\rho_E$ denotes the invariant phase-space volume on $\Gamma_E$ (i.e., the measure whose density relative to surface area is the inverse magnitude of the gradient of the Hamiltonian), and the ``macro-state'' $\Gamma_{y_0}$ is the following. Subdivide the $y$-axis into intervals of (very small) length $dy_0$, and let $\Gamma_{y_0}$ be the subset of $\Gamma_E$ for which $y(q,p)\in[y_0,y_0+dy_0]$. (In the following, we will not distinguish between $y$ and $y_0$.)
This entropy is shown in Appendix~\ref{app:ent} to be  a monotonic function of $y=y_0$, namely 
\begin{equation}\label{S352}
S=\frac{3N-5}{2}k\log y+ \mathrm{const.},
\end{equation}
where the added constant depends on $E$, $dy_0$, and $N$ but not $y$. Thus, the increase or decrease of $S$ coincides with that of $y$. 

As a consequence, entropy increases indeed with $|t-\tau|$ both towards the past and towards the future of the central time,\footnote{If  we modified the toy model by taking into account pure gravity (without any long range repulsive force arising, say, from a non-zero cosmological constant), then entropy would have to be defined differently. It would still increase, but by a rather different mechanism  (clustering in space and rather than expansion).} see Fig.~\ref{fig:curve}.
So, Fact 1 is true also in the toy model. That is, every solution of the dynamical laws of the toy model possesses an arrow of time (i.e., entropy increases monotonically) in the time interval $t>\tau$, and an opposite arrow of time in the interval $t<\tau$. In Carroll and Chen's cosmological model and in the model of Barbour et al., this is still true of the overwhelming majority of solutions. (In a universe governed by these models, the inhabitants in both intervals would, of course, perceive entropy as \emph{increasing} with time, as their mental arrow of time is aligned with the thermodynamic arrow.) Thus, the toy model illustrates how arrows of time can arise without a past hypothesis.\footnote{If we regard scale-invariant physical laws as defining a ``shape dynamics'' then we may wish to also define entropy in a scale-invariant way, contrary to what we did for the toy model. 
In this regard, see the analysis of Barbour et al.\ \cite{BKM13,BKM14,BKM15}.} 
We now turn to the conflict with Fact 2, which may even seem like a mathematical contradiction.

\section{Mathematical Contradiction}

For simplicity we take $E=N/2$.
Then for each solution, the image of the system's orbit in the $xy$-plane is of the form
\begin{equation}
y = x^2 + b\,,
\end{equation}
where $b\geq 0$ is the size of the system at the central time, see Fig.~\ref{fig:curve}. 

\begin{figure}[t]
	\centering
        \includegraphics[width=0.50\textwidth]{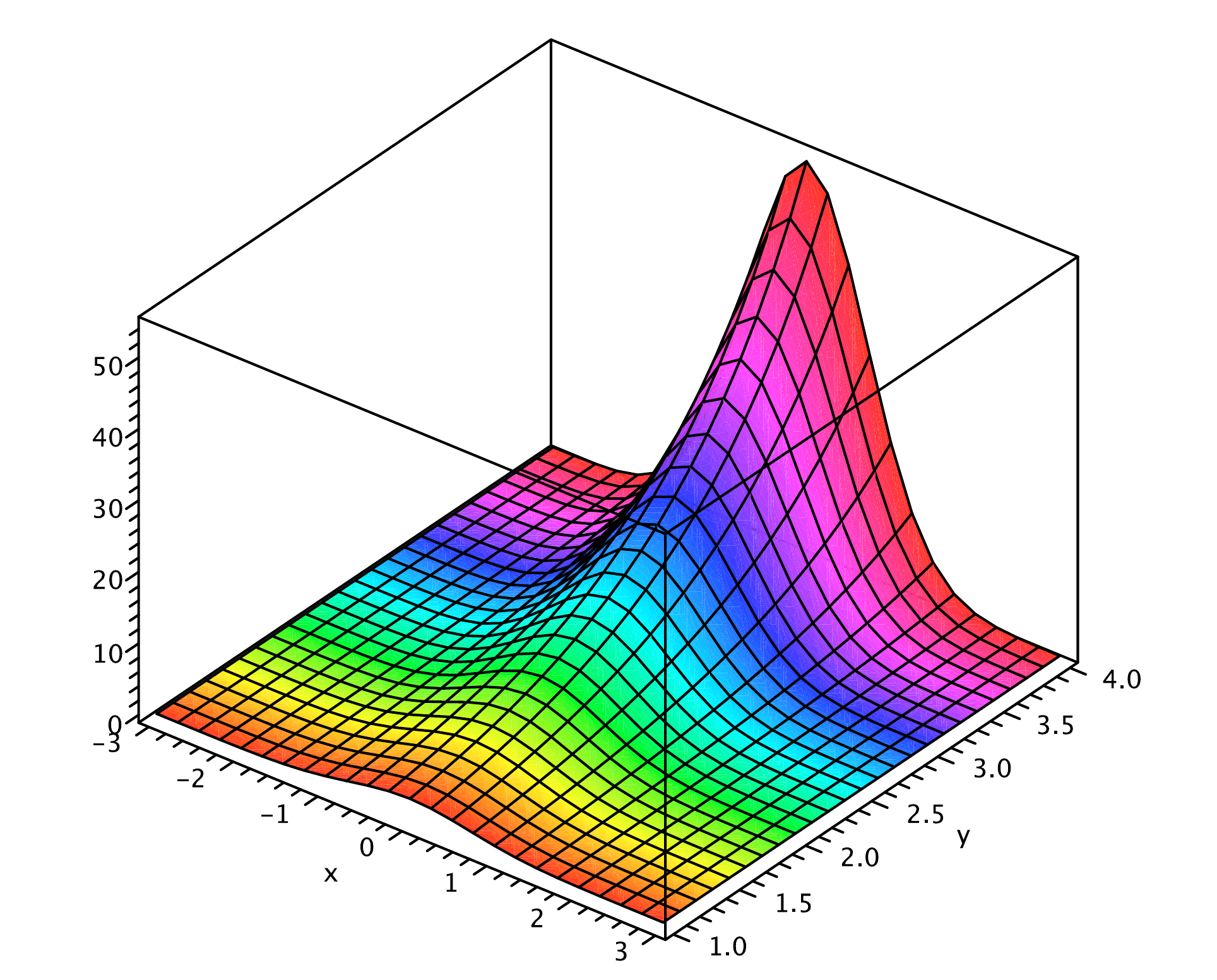}
	\caption{Plot of the the distribution $\rho(x,y) \sim e^{y-x^2}$.}
	\label{two}
\end{figure}

Let us now consider the  microcanonical (uniform) distribution $\rho_E$ on $\Gamma_E$. Its image on the $xy$-plane is (see Appendix~\ref{app:xy})
\begin{equation}\label{rhoexact}
\rho(x,y) \sim \Bigl( y - x^2 \Bigr)^{\frac{3}{2}(N-2)}
\end{equation}
restricted to the physical region
\begin{equation}\label{Lambdadef}
\Lambda= \{(x,y)\,|\,y\ge x^2\}\,.
\end{equation}
We shall consider instead the simpler density
\begin{equation} \label{rho}
\rho(x,y) \sim e^{y-x^2} 
\end{equation}
on the upper half-plane $\Lambda=\{(x,y)\,|\, y\geq 0\}$ (or on \eqref{Lambdadef}, that does not matter much), since it serves just as well for the points we wish to make; see Fig.~\ref{two} for a plot.
This distribution is (as it should be) constant along the orbits. 

Suppose that the system is in the macro-state $y= y_0$. Then  the conditional  probability distribution  is
\begin{equation}
\rho_{y_0} (x) =\rho (x |y= y_0)=  \frac{ \rho(x,y_0)} {\int_{-\infty}^{+\infty} \rho(x,y_0) dx } = \frac{1}{\sqrt{\pi}}  e^{-x^2}\,,
\end{equation}
and it is very likely that the system is near $x=0$ (say, $-2\leq x\leq 2$), that is, that it has near-minimal entropy. In other words, it is very likely that the system's entropy was not much lower in the past. Put yet another way, Fact 2 is true also in the toy model. Since $\rho_{y_0} (x)$ does not depend on $y_0$, by averaging over $y_0$  the same conclusion would seem to follow without any conditioning. More precisely, let $C$ be the vertical strip of width $4$ centered around  the $y$-axis. It is easy to see that  $\rho_{y_0}([-2,2]) > 0.9$  for all $y_0$, and one might be tempted to conclude that $\rho(C)/\rho(\Lambda) > 0.9$. (Of course, $\rho(C)=\infty=\rho(\Lambda)$.)

Now suppose we condition not on $y$ but on $b$, that is, not on horizontal lines, but on the parabolas $y= x^2 +b$, corresponding to the orbits of our system. Then, labeling the points on these parabolas by $x$, the  conditional distribution is $\rho_b (x) \propto 1$ (i.e., uniform). In particular, these conditional distributions assign small probability to $C$. Thus, reasoning as before we would seem to find that $\rho (C)/\rho(\Lambda) \ll 1$. (How small it would be  would depend on the size of the cutoff on $x$ which is required for the uniform distributions to be well defined.)

\begin{figure}[t]
    \centering
        \includegraphics[width=0.6\textwidth]{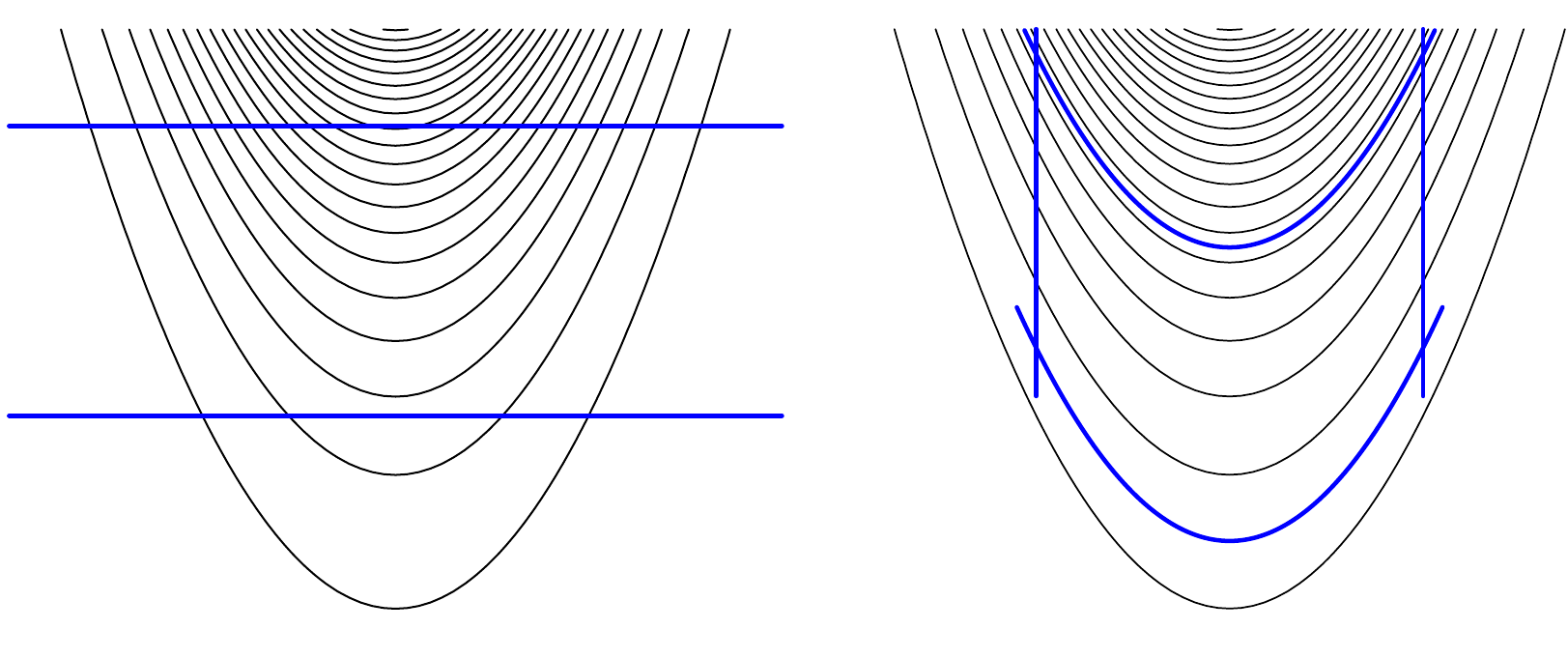}
    \caption{Left: The measure $\rho$ is normalizable when restricted to the region between two horizontal lines. For this cutoff, the central points have large weight.  Right: The measure $\rho$ is normalizable when restricted to the region delimited by two parabolas and   two vertical lines at $x=-L$ and $x=L$. For this cutoff, the central points have low weight for large $L$. }
    \label{one}
\end{figure}

We have arrived, it would seem, at a contradiction. 
 But the above reasoning would be justified only if  $\rho$ were normalized (or normalizable). For a probability (or normalizable) measure, the probability of an event is the average of its conditional probability with respect to a partition into ``fibers,'' and the total probability of the event of course does not depend on the choice of partition. But, as the previous example shows, the situation can be quite different for a non-normalizable measure.  Thus, because $\rho$ is non-normalizable, the above reasoning is not correct. (And if  we introduce cutoffs to make $\rho$  normalizable and then normalize it,  what we arrive at  for  $\rho (C)$ will depend significantly on the choice of  cutoff, see Fig.~\ref{one}.) The toy model thus illustrates how Fact 1 and Fact 2 can, indeed, be compatible.

\section{Should We Expect a Low Entropy  in the Past if We Make No Past Hypothesis?}

What seemed to be a blatant mathematical contradiction arising from the combination of Facts 1 and 2 has thus been resolved. But Fact 2 remains a fact, one that seems to imply that without the invocation of the past hypothesis we should expect the entropy of the universe to be higher in the past as well as in the future, both in the real world and in the ``central time models'' (i.e., the model of Carroll and Chen, that of Barbour et al., and the toy model). Indeed, should not the present macro-state (e.g., some $\Gamma_{y_0}$ in the toy model) be regarded as representing our best available evidence  about what is physically the case? If entropy increases towards the past, and not just towards the future, for the overwhelming majority of micro-states compatible with the evidence, should we not have an  overwhelming rational expectation that this bidirectional entropy increase in fact holds?

We believe that the answer is \emph{no, we should not}. Here is why.
Think about all the things for which the evidence corresponding to the present macro-state of the universe provides support. Among these things is our best physical theory itself. And it is only after we have such a theory that we can identify the set of micro-states that comprise the macro-state corresponding to our evidence. Indeed, the very nature of the phase space, some of whose subsets correspond to macro-states, depends on exactly which variables are the fundamental variables of the theory. Thus, some of the conclusions we  draw from the present evidence are pre-theoretical. Among these, in addition to the theory itself, are conclusions about the past, including perhaps the past hypothesis. And after we have our theory, and with it the detailed macro-states, what had been evidence for conclusions about the past remains so. 

Striking features of the present macrostate, such as the existence of
dinosaur  bones, and the fact that when we find a bone somewhere we usually find others nearby, are very strongly suggestive, providing evidence for the existence of  dinosaurs and, more generally,  that the earth and the universe had indeed  pasts like the ones we believe they had. The evidence we have in the present suggests  that the universe was more ordered in the past. This corresponds to a rather different assignment of weights than the uniform one for the microstates in, say, the present macrostate of the universe, one for which the overwhelming majority of micro-states for a uniform weighting is assigned negligible weight. Such an assignment is  in agreement with the past hypothesis, but is consistent with central time models as well.

\section{Types of Reasoning}

Let us reflect on the different types of reasoning involving probability that have been used in this paper: mathematical reasoning, evidential reasoning, and theoretical reasoning. The first involves just mathematics, and uses the mathematical structure provided by  probability theory as an important  mathematical tool. It should not be controversial. Nonetheless it can easily be conflated with evidential reasoning, for which probability is  also  an important ingredient. 

Evidential reasoning concerns the drawing of conclusions, with varying degrees of confidence, from some evidence. There are of course no clear-cut rules for how this should be done. One  procedure is suggested by the success of statistical mechanics. This procedure involves the use of the uniform distribution over all micro-states compatible with our present evidence to inform our conclusions about past behavior as well as future behavior. It is  responsible for the conviction that the invocation of the past hypothesis cannot be avoided. Insofar as the past is concerned, it is quite deficient. But with regard to the future it works extremely well.  That it does so is explained by statistical mechanics using the past hypothesis \cite{Alb}.

When we analyze the consequences of some theory, regardless of whether or not it involves the assumption of a past hypothesis, we take the theory as given, and are not concerned with the evidence for that theory or with how likely it is that the theory is correct. Rather we want to determine the extent to which the theory can account for the facts that the theory was intended to explain, facts such as the second law of thermodynamics. Such analyses we call theoretical reasoning. This, too, often involves the use of probability, but not merely as pure mathematics, and not really to provide a measure for different degrees of belief as appropriate to the evidence, but rather as a means of making more precise the extent to which what is to be explained holds for the overwhelming majority of space-time histories yielded by the theory.

We regard a physical phenomenon, such as entropy increase and irreversibility, as having been explained by a theory if we can show that it typically occurs for the  histories of the universe provided by that theory, i.e., that it occurs for the most by far of all theoretical histories. 
 If  the right behavior---for example, entropy increase---is established for {\em all} trajectories and not just for typical trajectories---as is the case for the central time models we are considering---who can ask for anything more? Thus, the models of Carroll and Chen and Barbour et al.\ show that it is not necessary to add any additional hypothesis to the laws of physics in order to explain our arrow of time.

\section{A Problem For Theoretical Reasoning?}

A variant of the puzzle we have discussed, of the tension between Fact 1 and Fact 2, exists also in the framework of theoretical reasoning: On the one hand, it naturally happens in typical solutions of the model of Carroll and Chen that galaxies, planetary systems, and life form over billions of years in very much the way they have actually formed in our universe. On the other hand, as a consequence of Fact 2, most phase points consistent with the present macro-state of the universe are near (say, within minutes of) their central times, suggesting that such macro-states most likely do not come into existence by a billion-year-long history. While the tension between these facts has a flavor very similar to that of our discussion above in terms of mathematical and evidential reasoning, it leads to further considerations that go beyond the scope of this paper and will be pursued in a subsequent work \cite{GTZ}. We argue there that this apparent paradox leads, upon closer analysis, not to a contradiction but rather to the conclusion that a theory of this kind without a past hypothesis can only work if entropy is unbounded from above (as it is in the models of Carroll and Chen and Barbour et al.).

\section{Conclusions}

The salient attractive trait of the models of Carroll and Chen and Barbour et al.\ is that they provide a possible explanation of the low entropy state that the universe was in at some point in the distant past, without just postulating, as the past hypothesis does, that the initial state of the universe was a very special one. In fact, in these models, most or all possible histories of the universe are such that the entropy curve $S(t)$ as a function of time $t$ first decreases (up to small fluctuations) from $+\infty$ to a minimum value and then increases back to $+\infty$, thus giving rise to a thermodynamic arrow of time in each of the two temporal regions. We have elucidated why this behavior may seem impossible in view of known facts of statistical mechanics, and why it is in fact possible. The relevant considerations had to do with the mathematical properties of non-normalizable measures that differ from those of normalizable ones, and with the conclusions that can validly be drawn about cosmological theories from the empirical fact that human beings exist, and similar facts.

\appendix

\section{Entropy Formula for the Toy Model}
\label{app:ent}

This appendix contains the calculation proving Equation~\eqref{S352}. Let us begin by repeating (a bit more generally) the definitions:
\be
\vq_i(t) = \vq_i(0)+t\vp_i(0)
\ee
\be
\vq_\cm = N^{-1} \sum_{i=1}^N \vq_i \,,\quad
\vp_\cm = \sum_{i=1}^N \vp_i
\ee
\be
y = N^{-1} \sum_{i=1}^N \Bigl( \vq_i-\vq_\cm \Bigr)^2
\ee
\be
\label{Edef}
E=\frac{1}{2} \sum_{i=1}^N \Bigl(\vp_i-N^{-1}\vp_\cm\Bigr)^2.
\ee
For later reference, we also note that
\begin{equation}\label{taudef}
\tau = -(2E)^{-1}\sum_{i=1}^N (\vq_i-\vq_\cm) \cdot (\vp_i-N^{-1}\vp_\cm). 
\end{equation}  
We write $n=3N$, $q=(\vq_1,\ldots,\vq_N)$, $p=(\vp_1,\ldots,\vp_N)$, and 
\be
U = \Bigl\{ (\vv_1,\ldots,\vv_N)\in\RRR^{3N}: \sum_{i=1}^N \vv_i=\boldsymbol{0} \Bigr\},
\ee
which is a subspace of dimension $n-3$.

Now we fix the value of $\vp_\cm$; for simplicity we consider $\vp_\cm = \boldsymbol{0}$ (which can be arranged by a Galilean transformation); that is, instead of $\RRR^n\times \RRR^n$ we consider the invariant subset $\RRR^n \times U$ as the phase space. The invariant measure on $\RRR^n \times U$ inherited from the volume measure on $\RRR^n\times \RRR^n$ is the $(2n-3)$-dimensional volume measure on $\RRR^n\times U$ (because $\vp_\cm(q,p)$ is a linear function on $\RRR^n\times \RRR^n$ so that, for each of its components, the norm of the gradient is constant). 

On $\RRR^n\times U$, $\vq_\cm(q,p)$ is a constant of motion because $d\vq_\cm/dt = N^{-1}\vp_\cm=\boldsymbol{0}$; we fix its value and can arrange by a translation that $\vq_\cm=\boldsymbol{0}$; that is, instead of $\RRR^n\times U$ we consider the invariant subset $U \times U$ as the phase space. The invariant measure on $U\times U$ inherited from the volume measure on $\RRR^n \times U$ is the $(2n-6)$-dimensional volume measure on $U\times U$.

On $U\times U$, we have that $E=\tfrac{1}{2}\sum_i \vp_i^2$, so the surface of constant energy is a sphere around the origin in momentum space; we fix the value of $E$ and can arrange by appropriate choice of units of length and time that $E=N/2$; that is, instead of $U \times U$ we consider the invariant subset $\Gamma_E:=U\times (U\cap \SSS_{\sqrt{N}})$ as the phase space, with $\SSS_r$ the sphere with radius $r$; it has the shape of a cylinder: an $(n-3)$-dimensional hyperplane times an $(n-4)$-dimensional sphere with radius $\sqrt{N}$. The invariant measure $\rho_E$ on $\Gamma_E$ inherited from the volume measure on $U\times U$ is the product of the $(n-3)$-dimensional volume measure on $U$ and the surface area measure on the $(n-4)$-dimensional sphere (because the norm of the gradient of $E$ is constant along $\Gamma_E$).

On $\Gamma_E$, we have that $y=N^{-1}\sum_i \vq_i^2$, so the surface of constant $y$ is an $(n-4)$-dimensional sphere around the origin in configuration space with radius $\sqrt{Ny}$ (times a $(n-4)$-dimensional sphere with radius $\sqrt{N}$ in momentum space). Therefore, the set
\be
\label{Gammasigma0}
\Gamma_{y_0} = \bigl\{(q,p)\in\Gamma_E:y_0\leq y(q,p)\leq y_0+dy_0\bigr\}
\ee 
is a spherical shell of thickness $dr=(N/2r)\, dy_0$ and radius $r=\sqrt{Ny_0}$, times a sphere of radius $\sqrt{N}$, and so
\begin{align}
\rho_E(\Gamma_{y_0}) 
&= (C\, r^{n-4} dr)(C\, N^{(n-4)/2})\\ 
&= \tfrac{1}{2} C^2 N^{n-7/2} y_0^{(n-5)/2} dy_0  
\end{align}
with $C$ the surface area of the $(n-4)$-dimensional unit sphere. After dropping the factor $dy_0$, this yields \eqref{S352}.

\section{Reduced Phase Space of the Toy Model}
\label{app:xy}

In this appendix we verify \eqref{rhoexact}. To this end, we need to compute $\rho_E(\Gamma_{x_0,y_0})$ with
\begin{align}
\Gamma_{x_0,y_0}= \bigl\{(q,p)\in\Gamma_E:
& x_0\leq x(q,p)\leq x_0+dx_0, \nonumber\\
& y_0\leq y(q,p)\leq y_0+dy_0  \bigr\}.
\end{align}
To this end, we use the fact that
\be\label{cupMp}
\Gamma_{x_0,y_0} = \bigcup_{p\in U\cap \SSS_{\sqrt{N}}} M_p\times\{p\}
\ee
with 
\begin{align}
M_p= \bigl\{ q\in U:
& \:x_0\leq x(q,p)\leq x_0+dx_0, \nonumber\\
& \:y_0\leq y(q,p)\leq y_0+dy_0 \bigr\}.
\end{align}
As we will show presently, $V_p := \mathrm{vol}_{n-3}(M_p)=V$ is independent of $p$. This fact, together with \eqref{cupMp}, implies that
\be
\rho_E(\Gamma_{x_0,y_0}) = V \,C \, N^{(n-4)/2}.
\ee
To compute $V_p$, we choose an orthonormal basis in $U$ whose first basis vector points in the direction of $p$ and introduce spherical coordinates for $q$ in this basis:
\be
q=\sqrt{Ny_0} \bigl( \cos \phi_1,\sin\phi_1\cos\phi_2,
\sin \phi_1\sin \phi_2\cos\phi_3,
\ldots, \sin \phi_1\cdots \sin \phi_{n-4}\bigr).
\ee
In these coordinates, by virtue of \eqref{taudef},
\be
x(q,p) = N^{-1/2} q_1 = \sqrt{y_0} \cos\phi_1\,,
\ee
so $M_p$ is a shell of thickness $dr=(N/2r)dy_0$ over a stratum of height $dq_1=\sqrt{N} \,dx_0$ of the $(n-4)$-dimensional sphere of radius $r=\sqrt{Ny_0}$; thus, $V_p$ equals $dr$ times the area $A_p$ of the stratum, $A_p=C' (r\sin\phi_1)^{n-5} \, r\, d\phi_1$ with $C'$ the area of the $(n-5)$-dimensional unit sphere; using $dq_1=r\sin \phi_1 \, d\phi_1$, we have that
\begin{align}
V_p 
&= C' (r^2-q_1^2)^{(n-6)/2} \, \sqrt{N} r \,dr \,dx_0\\
&= \tfrac{1}{2} C' N^{(n-3)/2} (y_0-x_0^2)^{(n-6)/2} \, dx_0\,dy_0\,,
\end{align}
indeed independently of $p$, and thus
\be
\rho_E(\Gamma_{x_0,y_0}) = \tfrac{1}{2} CC' N^{n-7/2} (y_0-x_0^2)^{(n-6)/2} \, dx_0\,dy_0\,,
\ee
which proves Equation \eqref{rhoexact}.

\bigskip

\noindent\textit{Acknowledgments.} 
We acknowledge support from the John Templeton Foundation [grant 37433 to S.G.\ and R.T.], the European Cooperation in Science and Technology [COST action MP1006 to N.Z.], and Istituto Nazionale di Fisica Nucleare [to N.Z.].

\end{document}